# Report on Earth Observation Missions and Ground Station Management using On-Demand Satellite Operation System

By Yuji S<small>AKAMOTO</small>[1]

[1]*Green Goals Initiative, Tohoku University, Sendai, Japan*

Since the launch of its first satellite in 2009, Tohoku University has continuously developed and operated Earth observation satellites and engineering demonstration satellites in the 50cm-class and CubeSat-class (up to 3U). The 50cm-class satellite launched into operation in 2021 enabled efficient operations through cloud-based management functions for both the satellite and ground stations, including automatic command generation. By 2022, up to eight operational satellites were simultaneously managed on a daily basis using three ground stations (Sendai, Hakodate, and Sweden). This paper presents the operational achievements to date and introduces the system that supports efficient satellite operations.

## 1. Introduction

In recent years, the constellation and utilization business of small satellites has been experiencing rapid growth worldwide. Since the launch of its first satellite in 2009, Tohoku University has continuously engaged in the development and operation of Earth observation satellites and engineering demonstration satellites in the 50cm-class and CubeSat-class (up to 3U). The mission sensors onboard the Earth observation satellites, as well as the observation infrastructure at the Hakodate and Kiruna stations, have been developed and operated in close collaboration with Hokkaido University.

In 2016, we published a paper on the construction of a low-cost ground station network and an operational system to support the operation of micro and nano satellites[1]. Since then, both the number of satellites under observation and the number of ground stations operating in real-time coordination have significantly increased. The 50cm-class satellite that began operations in 2021 enabled efficient management through cloud-based functions for both the satellite and ground stations, including automatic command generation. Additionally, a system was established to collect up to approximately 1 GB of image data per pass using a 20 Mbps X-band link. By 2022, up to eight operational satellites were simultaneously managed on a daily basis using three ground stations (Sendai, Hakodate, and Sweden). This paper presents the operational achievements to date and introduces the system developed to support efficient operations.

For reference, Fig. 1 shows the antenna facility at Tohoku University's primary satellite control station and past 50 cm-class remote sensing satellites. Detailed evaluations of the satellite systems and observation images of RISING-2, DIWATA-1, and DIWATA-2 are presented[2]-[5]. Matters closely related to DIWATA-2's operations, such as attitude control and onboard image recognition based on neural network techniques, are described[6][7].

This system serves as a platform for satellite operation management and satellite data management that can be utilized by multiple satellites, multiple ground stations, and multiple users (including non-space-sector operators), enabling "on-demand satellite operation." As long as the database server's capacity allows, the system is scalable. At the core of the web service is the satellite operation management (SOM), which provides communication opportunities between ground stations and satellites as well as imaging opportunities for target points requested by users, and distributes communication and imaging data online. Since immediacy from request to data delivery is required, full automation of both the satellite and ground station is a key objective.

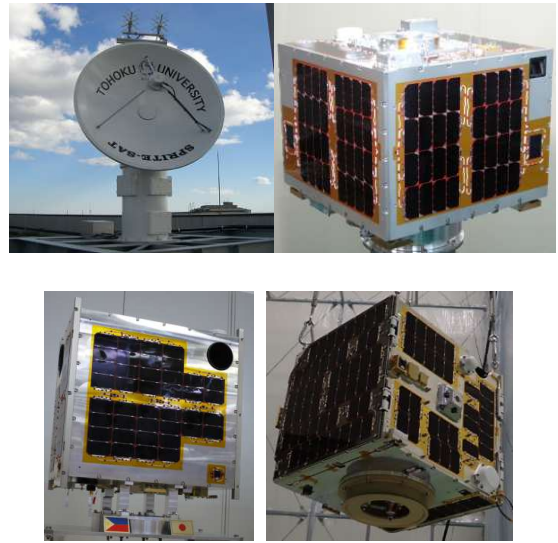

Fig. 1. Ground station CRESST operated by Tohoku University (upper left), and the 50cm-class satellites RISING-2 (upper right), DIWATA-1 (lower left), and DIWATA-2 (lower right). Note that DIWATA-1 and DIWATA-2 are primarily operated by DOST-ASTI or PhilSA, the national space agencies of the Philippines.

Planning of satellite operations is time-consuming and inevitably prone to human effort and error. Satellite operators must preliminarily investigate communication opportunities between ground and satellite, as well as times when target points can be imaged, and appropriately plan the command sequences to be uploaded to the satellite. This system allows for easy output of time lists for communication and imaging opportunities. It provides functions that automatically correct the execution times and target points of commands by combining predefined templates for each task.

Although complete automation within a unified system is difficult due to differences in planning methods specific to each satellite, this system can reduce up to 90% of the preparation time. Operation planners can concentrate their effort on final checks of satellite-specific settings (such as power and memory management), which helps reduce human error. Additionally, there will be room to devote resources to developing software that automates this final verification step. Detailed explanations of SOM and the related satellite data management function SDM (Satellite Data Management) are described[8][9].

The satellite list includes those for which CRESST has served as either the primary or secondary ground station. As of May 2025, the total number is 13 satellites. The start and end months of operations are indicated by S(month) and E(month), respectively. The communication targets include 50cm-class microsatellites (M), 2U CubeSats (2U), and 3U CubeSats (3U).

SPRITE-SAT was the first satellite developed by Tohoku University, but it experienced a critical malfunction in its main function within approximately two weeks. Nevertheless, signal monitoring continued for 15 years until operations concluded in 2024. RISING-2 remained operational but was decommissioned in its tenth year as further observations became impractical. Currently, only DIWATA-2 and HIBARI remain in operation. Three new observation targets are planned to be added in 2025–2026. Note: RISING-4 is a development code name and differs from its official designation in two line element (TLE) records and other sources.

CRESST is equipped with U-band command, S-band command, S-band telemetry (up to ~100 kbps), and X-band telemetry (up to ~20 Mbps) capabilities. HU-Kiruna supports S-band telemetry (up to ~1 Mbps), and HU-Hakodate supports X-band telemetry (up to ~20 Mbps).

RISING-2 was the first satellite in this development group to successfully acquire meaningful multi-wavelength images with a ground resolution of 5 meters. It was equipped with a U-band command and S-band telemetry (100 kbps) system. DIWATA-2 is equipped with the SMI+ERC (Space-borne Multispectral Imager + Enhanced Resolution Camera) sensor system, which provides multi-wavelength imaging at 54-meter resolution, and with the HPT (High Precision Telescope) camera for high-resolution imaging (4.7 meters, RGBN: red, green, blue, and near-infrared). The addition of an X-band telemetry system (2.4 Mbps) enabled an increase in data download volume. RISING-4 further enhanced performance with an SMI sensor offering 47-meter resolution, an HPT camera with 2.2-meter resolution (RGBN), and a 20 Mbps X-band telemetry link. Unfortunately, due to its orbit resulting from deployment from the International Space Station (initial altitude ~420 km), the satellite decayed in 2 years and 1 month. Among the remote sensing satellite series described in this study, only DIWATA-2 remains operational as of May 2025.

| Year | GRS | | | Satellites(type**) | | | | | | | | | |
|---|---|---|---|---|---|---|---|---|---|---|---|---|---|
| | CRESST | HU-Kiruna | | SPRITE-SAT(M) | | | | | | | | | |
| 2009 | S(1)* | S(3) | | S(1) | | | | | | | | | |
| 2010 | \| | / | | / | | | | | | | | | |
| 2011 | \| | / | | / | RAIKO(2U) | | | | | | | | |
| 2012 | \| | / | | / | S(10) | | | | | | | | |
| 2013 | \| | / | | / | E(8) | RISING-2(M) | | | | | | | |
| 2014 | \| | / | | / | | S(7) | S-CUBE(3U) | | | | | | |
| 2015 | \| | / | | / | | \| | S(9) | DIWATA-1(M) | | | | | |
| 2016 | \| | / | | / | | \| | E(11) | S(4) | | | | | |
| 2017 | \| | / | HU-Hakodate | / | | \| | | | DIWATA-2(M) | ALE-1(M) | | | |
| 2018 | \| | / | S(12) | / | | \| | | | S(10) | RISESAT(M) | ALE-2(M) | | |
| 2019 | \| | / | \| | / | | \| | | | S(1) | S(1) | S(12) | ASTERISC(3U) | |
| 2020 | \| | / | \| | / | | E(4) | | | \| | \| | \| | RISING-4(M)*** | HIBARI(M) |
| 2021 | add X | \| | \| | / | | | | | \| | \| | \| | S(3) | S(11) | S(11) | IHI-SAT(3U) |
| 2022 | \| | \| | \| | / | | | | | \| | \| | \| | \| | \| | \| | S(3)-E(11) |
| 2023 | \| | \| | \| | / | | | | | E(3) | E(10) | E(4) | E(4) | \| | \| |
| 2024 | \| | \| | E(1) | E(7) | | | | | \| | | | \| | \| |
| 2025 | \| | \| | | | | | | | | | | | E(1) | \| |

\* S(month)=Start, E(month)=End  
\*\* (M)=50cm microsat, (2U)=2U CubeSat, (3U)=3U CubeSat  
\*\*\* internal name, different from the official designation  
"/" means limited or degraded operations

Fig. 2. Satellite operation history by Tohoku University and its affiliated GRS (as of May 2025), Note: The ground stations and satellites are owned by various institutions, many of which are outside of Tohoku University.

As relatively recent examples, downlink data volumes from five university satellites launched since 2018 are shown in Fig. 4. Among the satellites, RISING-4—equipped with a 20 Mbps X-band transmitter—achieved the largest cumulative downlink volume at 171 GB over an operational period of 2 years and 1 month. Among the ground stations, CRESST (X-band) recorded the largest volume at 131 GB. HU-Kiruna, while using S-band, also achieved a large cumulative volume due to its location in the polar region enabling frequent passes of sun-synchronous satellites, and because of its higher link margin (compared to CRESST), which allowed operation at up to 1 Mbps.

## 3. Operations Using Satellite Operation Management (SOM)

This section introduces the Satellite Operation Management (SOM) function. By centrally managing each ground station through SOM, the ground station can operate automatically in sync with satellite pass times, and the collected data is synchronized to the cloud, allowing operators to check data volumes and other information via mobile devices. In the case of the RISING-4 satellite, SOM contributed to improved operational efficiency through its command list auto-generation support function, resulting in the highest-ever download volume, mainly consisting of images. This coordination function is expected to be essential for upcoming 50cm-class remote sensing satellites as well.

### 3.1. Features of SOM

The proposed system is a platform for satellite operation management and satellite data management that can be utilized by multiple satellites, ground stations, and users, enabling "on-demand satellite operation". SOM is implemented as a web service with an API interface connected to a database server (hereafter referred to as the SOM server) that can be managed via SQL. In response to user requests, it provides communication opportunities between ground stations and satellites, imaging opportunities by satellites for target locations, and delivers communication and imaging data online. As shown in Fig. 5, the proposed system offers three main functions: (1) ground station sharing, (2) satellite operation planning support, and (3) satellite image distribution. Multiple satellites and ground stations can be accessed by both satellite operators and data users through this system.

The system supports the automatic creation of satellite operation plans and enables on-demand imaging services for

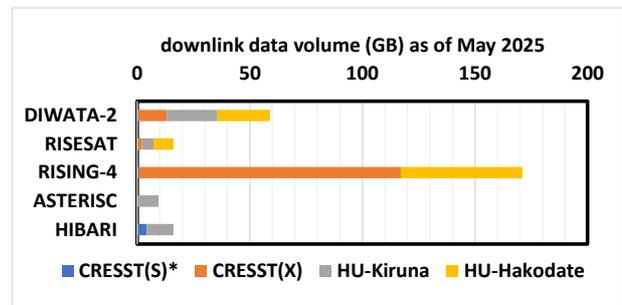

| Link* | GRS | | Satellites | | |
|---|---|---|---|---|---|
| | CRESST | HU-Kiruna | RISING-2 | DIWATA-2 | RISING-4 |
| U-CMD | 1.2kbps | No | Yes | Yes | No |
| S-CMD | 1kbps | No | No | Yes | Yes |
| S-TLM | 100kbps | 1Mbps | Yes(100K) | Yes(1M) | Yes(1M) |
| X-TLM | 20Mbps | No | No | Yes(2.4M) | Yes(20M) |
| multispectral imager | | | (**) | 54m GSD | 47m GSD |
| high-precision telescope | | | 5m GSD | 4.7m GSD | 2.2m GSD |

* U=UHF, CMD=command uplink, TLM=telemetry downlink
(**) no dedicated imager, but the telescope has multispectral capability

Fig. 3. Evolution of satellite payload capabilities and telemetry data rates

| satname | downlink data volume (GB) | | | | total(/sat) |
|---|---|---|---|---|---|
| | CRESST(S)* | CRESST(X) | HU-Kiruna | HU-Hakodate | |
| DIWATA-2 | 0.5 | 12.7 | 22.5 | 23.3 | 58.9 |
| RISESAT | 0.1 | 2.0 | 5.1 | 8.9 | 16.1 |
| RISING-4 | 0.7 | 116.4 | NA | 54.0 | 171.1 |
| ASTERISC | 0.2 | NA | 9.2 | NA | 9.4 |
| HIBARI | 4.2 | NA | 11.9 | NA | 16.1 |
| total(/GRS) | 5.7 | 131.1 | 48.7 | 86.2 | |

* Only CRESST(S) shows the cumulative data volume since Jan. 2022

Fig. 4. Actual downlink data volumes by satellites and ground stations

data users. A key feature is that imaging requests from data users can be immediately reflected in the operation plans prepared by satellite operators. The satellite operation planning management function significantly reduces the burden of operation preparation, especially for new satellite operators. Planning satellite operations is time-consuming, and it inevitably involves human effort and the risk of human error. Satellite operators must investigate in advance the communication opportunities between ground and satellite, as well as the possible imaging times for target locations, and appropriately plan the command sequences to be uploaded to the satellite.

During the satellite's pass over a ground station (less than 15 minutes for satellites below 650 km altitude), command data is uploaded in bulk. These command sequences define actions issued at either absolute or relative times. In planning, factors such as satellite attitude control, power management, and onboard memory management must be considered. In practical operations, sending commands once per day per satellite is the standard practice for satellites under our operation.

However, generating the command data to be transmitted requires significant time. In initial operations in particular, time is critical, and often there is not enough time for multiple members to coordinate; thus, a single person may bear a heavy planning burden. If there is a flaw in the planned commands, it could result in a critical failure of the satellite's fundamental functions.

This system allows for easy output of time lists for communication and imaging opportunities. It provides functions to automatically adjust execution times and target coordinates by combining predefined command templates for each task. Operation planners can then focus their efforts on final checks for satellite-specific settings (e.g., power and memory management), which helps reduce human error. In addition, it provides room to allocate resources for developing software that automates this final verification step.

Operation plans for satellites and ground stations are managed in units called "sessions" within the database. A single session, typically lasting 10 to 30 minutes, includes a group of tasks such as attitude maneuvers, Earth imaging (or data communication), and device suspension actions. For example, command uploading, Earth imaging, and data downloading would constitute three separate sessions. The database tables correspond to each abstraction model: Satellites, Sensors, CommLocations, CapLocations, Requests, and Sessions.

Operation of ground stations and satellites using the support system SOM began in March 2021. Based on actual operational experience, the system concept has been concretely refined.

### 3.2. Management method for satellite operation planning

An operation plan consists of a set of operation sessions. Within a single session, the satellite powers on the relevant equipment, performs attitude control (targeting) toward a ground station or an imaging location, carries out data download or new image acquisition, and finally powers off any unnecessary equipment to return to an idle state, as illustrated in Fig. 6. Fig. 7 shows the sequence of sessions included in a single command upload.

The pass times over ground stations or imaging locations can be simply calculated from the satellite's orbital elements and the coordinates of the locations. Based on the pass time (i.e., the time of maximum elevation), the system determines the start and end times of each session by assigning predefined lead and lag times for each session type. Satellite communication and imaging operations are determined by the uploaded command sequences. Therefore, no new actions can be inserted during periods when the satellite is out of contact with the ground station. If a satellite can be observed once every 12 hours by a single ground station, up to 12 hours of waiting may be necessary before image acquisition. Assuming that the finalization of the command list takes about one hour, data users can submit imaging requests via Web API to target observations between one and twelve hours into the future.

There are two types of sessions: communication sessions and imaging sessions. In principle, satellite operators plan communication sessions, while data users submit requests for imaging sessions. Users select desired items from a predefined list of request templates. The system then calculates the start and end times for each session and presents a tentative session (Ses-A) to the user. Upon confirmation of the selected items, the system registers a confirmed session (Ses-B) in the server. Users may edit the associated models—such as the satellite, sensor, ground station location, and imaging target. These model definitions and session creation functions are provided within the SOM system, shown in Fig. 8.

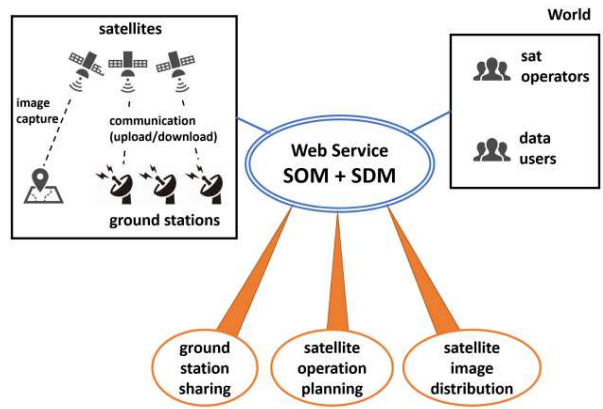

Fig. 5. Concept of an on-demand satellite operation system

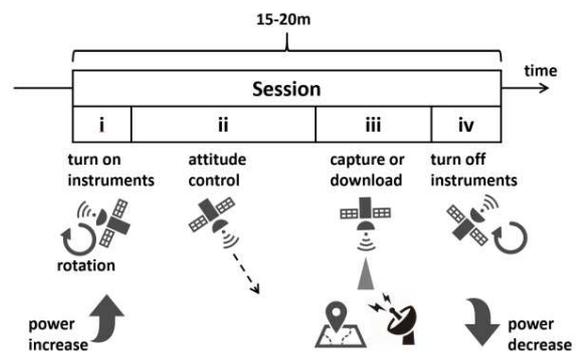

Fig. 6. Example of satellite operations within a single session

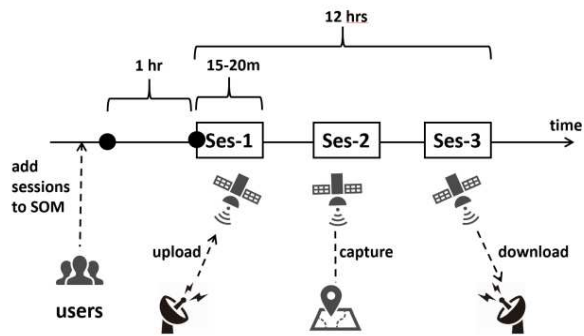

Fig. 7. Sequence of sessions from command upload to data download

Satellite operators then refer to the Ses-B list to finalize the command list to be uploaded to the satellite. This system automatically merges the command templates corresponding to each session, updates the session start times and target coordinates, and generates a complete command file. Operators can then focus on satellite-specific final adjustments, significantly reducing total workload, as shown in Fig. 9.

Communication sessions for all satellites and all ground stations are automatically registered (and updated) every day for the next seven days. During this process, interference between observation passes is checked, and final priority is determined based on priority level and maximum elevation angle (MEL). Each ground station selectively extracts pass information from the database for its own visible sessions. Based on that information, the system performs automated operations including antenna control from Acquisition of Signal (AOS) to Loss of Signal (LOS), and synchronization of downlinked files after LOS. Antenna angles for each time step are calculated locally.

Attributes of communication sessions include: satellite name, ground station name, communication link name, maximum elevation angle, whether the satellite is sunlit, whether there is interference with other communication passes, enable/disable flag, representative timestamps (AOS, MEL, LOS), and communication priority. Each communication pass corresponds to one record.

Attributes of imaging sessions include: satellite name, imaging sensor name, target location (name and coordinates), closest approach time, satellite elevation angle at the target, satellite roll angle, whether the satellite is sunlit, solar elevation angle at the location, orbital plane solar elevation angle, local time, resolution factor (1.0 for nadir, 2.0 at twice the swath width), and forecasted cloud cover.

Example of a session list in SOM database is shown in Fig. 10. Communication (Comm) and Imaging (Capture) sessions are managed in a unified table. Both types of sessions can be listed together and sorted chronologically.

### 3.3. Case study

This section presents a case study illustrating how SOM is utilized in daily satellite operations and ground station management. The example is based on the command generation support actually used in the operation of the RISING-4 satellite.

**(1) Satellite specifications and imaging method**

This case considers the use of a 50cm-class microsatellite, equipped only with body-mounted solar cells, for Earth observation. Since the satellite does not have deployable solar panels, its power resources are limited. Continuous Earth observation is difficult, and imaging is therefore limited to specific designated areas. On the other hand, the lack of deployable mechanisms significantly reduces development time, and it has also been observed to help extend the operational life of onboard equipment. By omitting deployable panels, the satellite reduces atmospheric drag, thereby helping to prevent rapid orbital decay. A significant drop in altitude can disrupt the sun-synchronous orbit, which would severely impact the observation mission. This low-power, on-demand operation style is suitable for missions focusing on observations over specific countries or regions.

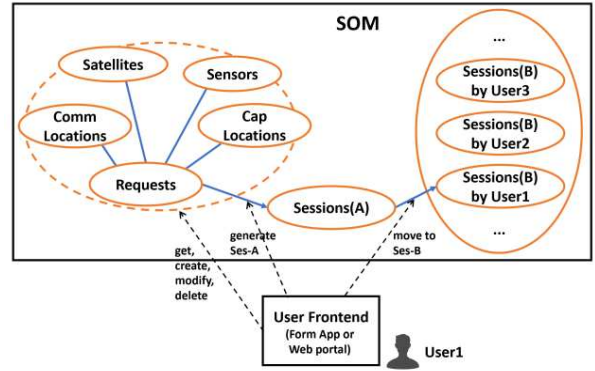

Fig. 8. Relationships among models comprising SOM

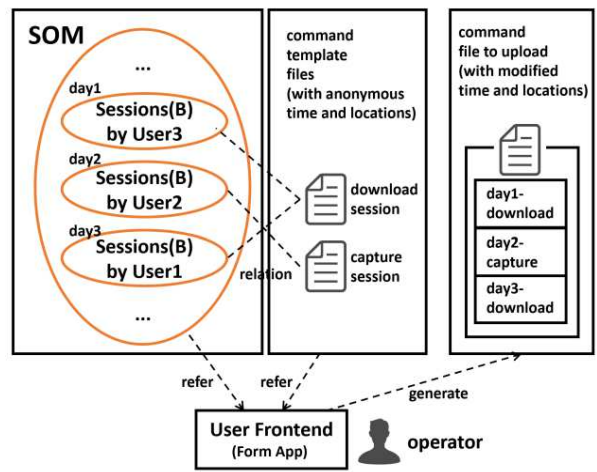

Fig. 9. Command file generation based on session list

Fig. 10. Example of session list in SOM database

A drawback of the on-demand approach is that each observation requires a series of operations: powering on devices, attitude determination, maneuvering, target orientation and imaging (or communication), and transition to standby. Consequently, the command planning becomes complex. Once a standard procedure is established, certain parameters must be updated for each session, such as:
a) The absolute start time of the session (subsequent commands follow relative timing)
b) Target coordinates for orientation
c) Time of closest approach (MEL time) to the target, i.e., imaging time

In practice, more complex conditional branching is required. For example, different combinations of attitude sensors are used depending on whether the satellite is in sunlight or eclipse, and some sensors need to be powered on before entering eclipse. Star trackers must avoid the Earth, Sun, and Moon when acquiring star fields. The satellite selects a preferred orientation using if-statements based on the orbital plane solar elevation angle and local time.

The satellite carries the following four types of cameras:
- HPT (High Precision Telescope): 2.2 m resolution, 3.6 × 2.7 km image size, RGBN
- SMI (Spaceborne Multispectral Imager): 47 m resolution, 77 × 58 km image size, selectable wavelength from 430–1020 nm in 1 nm increments
- MFC (Middle Field Camera): 35 m resolution, 58 × 44 km image size
- WFC (Wide Field Camera): 180° × 134° field of view

In nominal operation, imaging is typically conducted using either the SMI+MFC or HPT+MFC combination. The conditions for image acquisition are assumed to be as follows:
- Nadir attitude inclination during imaging: must not exceed 1.3 × the camera's ground resolution (in meters)
- Solar elevation angle at the target location: ≥ 5° for HPT, ≥ 25° for SMI
- Forecasted cloud cover: ≤ 25%

For communication, the satellite uses both S-band and X-band: the S-band for command uplink and attitude log replay, and the X-band for downlinking image data.

**(2) Operation planning**

The following procedure is executed every two days for each satellite:
- Register and adjust sessions (communication or imaging) in the database
- Based on the imaging sessions, register the onboard memory addresses for data storage (automatically added/removed)
- On the client application, execute the command (CMD) generation process with a button click to retrieve the CMD file

Predefined template files corresponding to communication and imaging sessions are required. The software automatically modifies the command content based on session information (e.g., timing) and memory references (e.g., address). Operators upload the CMD file to the satellite via the S-band link.

In step 1, communication sessions are pre-registered by the system (with "disabled" as the default), and the operator manually enables them.

For imaging sessions, a list of candidate imaging opportunities across all locations is generated in response to a single request, based on standardized conditions (e.g.,

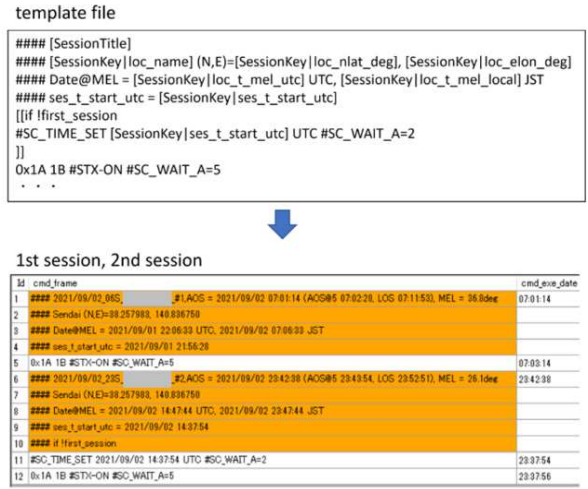

Fig. 11. Example of template file and resulting CMD file

allowable cloud cover). Final selection is performed manually.

The CMD file generated in step 3 includes:
- One command per line, consisting of a hex command code + command name (as comment) + wait time in seconds
- An absolute time wait command inserted at the beginning or mid-sequence
- The absolute time is automatically replaced based on session information

The final data format sent to the satellite differs from this intermediate CMD file and is converted using a separate utility.

**(3) Command generation support**

The CMD generation process references predefined template files. Fig. 11 shows examples of templates and their conversion into actual CMD files. In actual operations, four types of template files are defined:
- Communication sessions: routine satellite management and attitude log replay
- Communication sessions: X-band data download
- Imaging sessions: SMI+MFC imaging
- Imaging sessions: HPT+MFC imaging

In type 1, routine management tasks are defined for the primary control station. The first session in each CMD file corresponds to the command upload pass and executes tasks such as orbital element (TLE) updates and magnetic detumbling control. Subsequent sessions replay attitude logs. The attitude log replay task includes the following sub-steps:
a) Absolute time wait
b) If in eclipse: power on FOG gyros before eclipse entry, initialize attitude using Sun/magnetic sensors, and perform gyro integration during eclipse
c) Start 2-axis attitude control using wheels (ignoring yaw axis to conserve power)
d) Replay attitude logs
e) Stop attitude control using wheels

In each template, replaceable keywords (e.g., ses_t_start_utc for session start time) are used for session-specific values. Time variables allow minor adjustments using + or − seconds. A custom syntax allows if-statements for conditional command selection.

In type 2, the session flow is similar to type 1, with the following differences and additions (LOS5 refers to the time when satellite elevation drops to 5° before LOS):

b+) Power on SHU (science handling unit, science computer) and XTX (X-band transmitter)

d) Replay image data (loop until LOS5)

e+) Power off SHU and XTX

Step d) loops through playback addresses until LOS5. A while statement is used to implement this loop. If no data remains, the loop ends before LOS5.

In types 3 and 4, the session flow also follows type 1 with the following differences and additions:

b+) Power on SHU and relevant sensors (SMI, HPT, etc.)

c) Start 3-axis attitude control using wheels. First, set orientation for star tracker (STT) measurement in roll-pitch-yaw (RPY) angles. Then switch to target location for imaging

d) Execute image capture, store attitude logs, and save image data

e+) Power off SHU and sensors

In step c), the STT orientation is selected using conditional branches based on orb_sun_deg (orbital plane solar angle) and sat_t_mel_lst (local solar time at MEL).

## 4. Conclusion

This paper reviewed the continuous development and operation of Earth observation satellites and engineering demonstration satellites in the 50cm-class and CubeSat-class (up to 3U), and introduced the system developed to support efficient satellite operations based on actual operational achievements.

Regarding command generation, once the templates are finalized during the initial operation phase, the system automatically replaces unspecified parameters in the templates with actual values, such as observation times (or communication times) and observation targets (d target coordinates and imaging timing), for each operation.

Due to the complex conditional branching involved, the templates tend to resemble source code with extensive use of if and while structures—an aspect that still offers room for improvement. Additionally, when multiple satellites share the same communication pass, or when observation targets overlap in time, the final decisions are currently made manually by operators. Automation of these processes remains an area requiring further enhancement.

Moving forward, we aim to continue refining the system by leveraging the accumulated knowledge and practical data obtained through actual satellite operations.

### bibliographic note

An earlier version of this work was presented at the 35th International Symposium on Space Technology and Science, and 14th Nano-Satellite Symposium (ISTS 35th, July 12–18, 2025).

This paper is a revised and extended version prepared in January 2026.

**Appendix: Selected Observation Images Obtained Using the On-Demand Satellite Operation System**

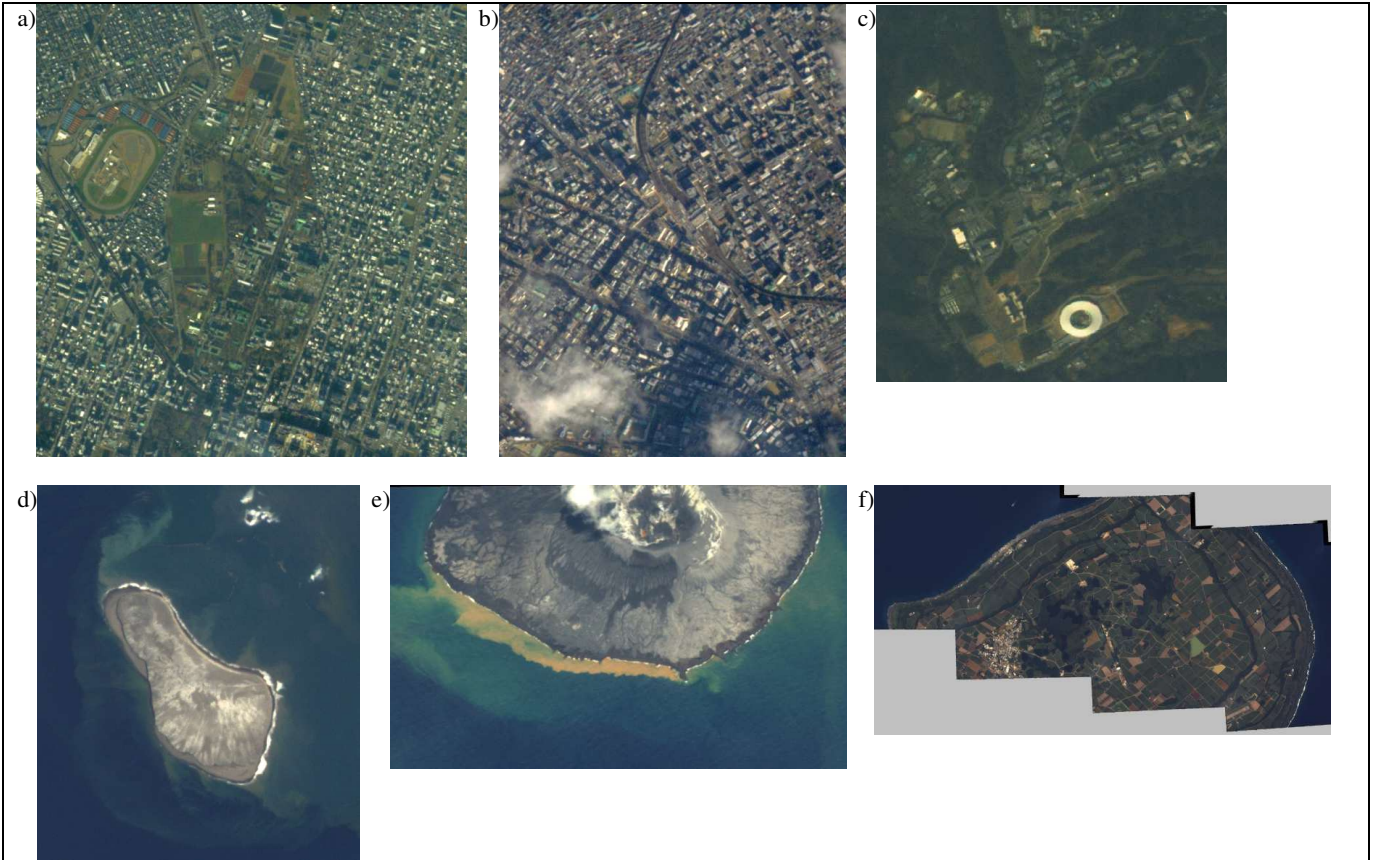

Fig. A-1   Selected Images by High Precision Telescope (HPT): a) Sapporo Station and Hokkaido University, b) Sendai Station, c) Tohoku University Aobayama Campus, d) Fukutoku-Okanoba (approx. 1 km in diameter), e) Nishinoshima (4.4 km$^2$), f) Minami-Daito Island.

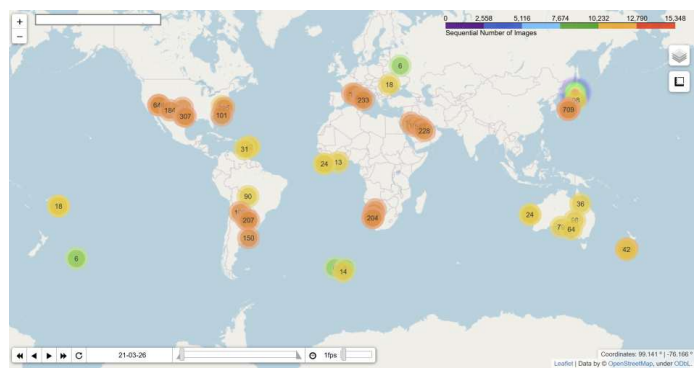

Fig. A-2   Example of a satellite image distribution portal map

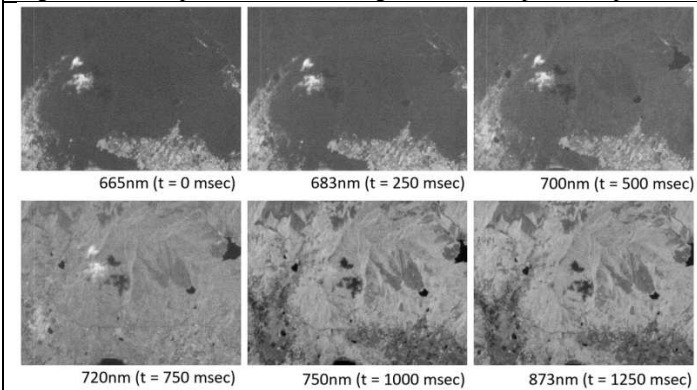

Fig. A-3   Images by Spaceborne Multispectral Imager (SMI)